\documentclass[12pt]{article}
\usepackage{cite,psfig}

\def\be{\begin{equation}}
\def\ee{\end{equation}}
\def\bea{\begin{eqnarray}}
\def\eea{\end{eqnarray}}
\def\msb{\overline{\mbox{MS}}}
\def\mmsb{\overline{\mbox{\scriptsize MS}}}
\newcommand{\mst}[2]{\mbox{\raisebox{-1mm}{$\,\stackrel{#1}{\scriptstyle 
#2}\,$}}}

\begin{document}

\thispagestyle{empty}
\vspace*{.5cm}
\noindent
{\large HD-THEP-99-42\hfill September 1999}\\
{\large HD-TVP-99-9}
\vspace*{2.5cm}

\begin{center}
{\Large\bf The Semiclassical Gluon Distribution\\[.3cm] 
at Next-to-Leading Order}
\\[2.5cm]
{\large H.G. Dosch, A. Hebecker, A. Metz and H.J. Pirner}\\[.5cm]
{\it Institut f\"ur Theoretische Physik der Universit\"at Heidelberg\\
Philosophenweg 16} \& {\it 19, D-69120 Heidelberg, Germany}\\[2.1cm]

{\bf Abstract}\end{center}
\noindent
The interaction of the partonic fluctuation of the virtual photon in deep 
inelastic scattering with soft color fields describing the hadron is 
treated in an eikonal approximation. It is known that, in this approach, 
the small-$x$ limit of the leading-order gluon distribution $xg(x,Q^2)$ is 
a constant characterizing the averaged local field strength in the target.
Matching the next-to-leading order calculation in this semiclassical 
framework with the one-loop parton model result, we obtain the 
next-to-leading order contribution to $xg(x,Q^2)$. It shows a ln$(1/x)$ 
enhancement at small $x$ and is sensitive to the large distance structure 
of the target. The final expression is a simple integral over non-Abelian 
eikonal factors measuring the target color field. We derive a quantitative 
relation between the short-distance cutoff of this integral and the scale of 
the gluon distribution function in the $\msb$ scheme. Our calculation 
demonstrates that higher order contributions can be systematically included 
in the semiclassical approach. 
\vspace*{2cm}
\newpage

%%%%%%%%%%%%%%%%%%%%%%%%%%%%%%%%%%%%%%%%%%%%%%%%%%%%%%%%%%%%%%%%%%%%%%%%%%%%
\section{Introduction}
The small-$x$ limit of deep inelastic scattering (DIS) remains one of 
the most challenging problems on the interface of perturbative and 
non-perturbative QCD. It can be expected that both the perturbative physics 
described by QCD resummation techniques~\cite{dglap,bfkl} and the 
non-perturbative soft dynamics underlying the growth of total hadronic 
cross sections~\cite{dl} are important for a complete picture of the 
small-$x$ limit of structure functions. In this paper, we systematically 
relate models for the soft color fields, which we consider to be a 
promising tool for the characterization of the hadron, to the hadronic 
gluon distribution, which is the basic object of the perturbative treatment. 

We are interested in a region of $x$ and $Q^2$ where, although $x$ is very 
small, the concept of parton distributions and the conventional DGLAP 
evolution equations~\cite{dglap} are applicable. For the description of 
the DIS process, the gluon distribution $g(x,\mu^2)$, which dominates in the 
small-$x$ region, has to be given at some scale $\mu^2<Q^2$. Thus, the 
problem of the non-perturbative structure of the target hadron has to be 
addressed. 

Our basic starting point is the idea of describing high-energy processes in 
QCD by studying the eikonalized interaction of energetic partons with 
soft color fields~\cite{nac,dk}. In this approach, 
which has been used for the description of DIS in~\cite{dgp}, soft color 
fields in the center-of-mass frame of the collision mediate the interaction 
between the energetic projectile and target quarks. Progress towards the 
description of the energy dependence of cross sections in the framework of 
soft color field dynamics was reported, e.g., in~\cite{sc}. 

A closely related approach to both inclusive and diffractive DIS, which 
treats the target proton as a superposition of soft color fields, was 
developed in~\cite{bdh,bgh}. In fact, this semiclassical approach 
reproduces the treatment of~\cite{nac,dk,dgp} if the dynamics of the 
underlying color field is modelled on the basis of the stochastic 
vacuum~\cite{ds} and a phenomenological ansatz for the constituent-quark 
wave function of the proton is made. For different approaches to the 
description of the target color fields see, e.g.,~\cite{hks}. However, in 
the following we are completely general and use no model-specific features 
of the target color field configurations or, equivalently, the wave 
functional of the proton. 

In the present paper, the gluon distribution $g(x,\mu^2)$ at $x\ll 1$ and 
$\mu^2\gg\Lambda^2$ (where $\Lambda$ is a soft hadronic scale) is 
calculated for a target given by soft color field configurations. Here 
`soft' means that all momentum components of the field are ${\cal O} 
(\Lambda)$. Following~\cite{mue}, a scalar `photon', coupled 
directly to the gluon field, is used as a convenient theoretical tool for 
extracting the gluon distribution. The leading order calculation gives a 
constant for $x\,g(x,\mu^2)$, which is a measure for the averaged gluon 
field strength in the target. This is in agreement with the seminal paper 
of Mueller~\cite{mue} and with~\cite{bgh}, the spirit of which we follow 
closely. 

In our opinion, it is crucial, both from a theoretical and a 
phenomenological perspective, to demonstrate the viability of the 
approach at higher orders. However, already at next-to-leading order the 
gluon distribution is scheme dependent and a careful matching of the 
partonic calculation (we use the $\msb$ scheme) and the semiclassical 
calculation is necessary to obtain an unambiguous result. With this result, 
we treat problems that were not addressed in the one-loop calculations 
of~\cite{mue,mue1}, where the scheme dependence was not discussed. In fact, 
the problems of regularization and scheme dependence arise immediately if 
one attempts to translate the one-loop, unintegrated gluon distribution 
of~\cite{mue1} into a correction to the leading term in the spirit 
of~\cite{bgh} (see~\cite{b} for a comparison of the results of~\cite{bgh} 
and~\cite{mue1} in the case of the quark distribution). As emphasized 
in~\cite{bgh}, where diffractive and inclusive quark and gluon 
distributions were calculated in the semiclassical approach, the inclusive 
gluon distribution dominates the small-$x$ region. Therefore, we expect 
that our next-to-leading order result is the dominant correction relevant 
as input for the next-to-leading order DGLAP evolution.
%for future analyses in this framework. 

In our approach, the most intricate part is the semiclassical calculation 
at next-to-leading order. Working in Feynman gauge, we employ 
the optical theorem and calculate the forward scattering amplitude. In the 
high-energy limit, certain diagrams can be dropped. The remaining 
contribution is given in the form of a two-gluon production cross 
section. In this way, the identification with the parton model result 
becomes simple since the dangerous high-mass region, where the 
semiclassical approximation fails, cancels explicitly. 

Let us note that, for a soft hadron governed by the single scale $\Lambda$, 
the perturbative expansion of the gluon distribution makes no sense. 
However, following ideas of~\cite{mv} (see also the recent calculations 
of~\cite{kov}), we can always assume that we are dealing with a very large 
target, in which case the gluon distribution becomes calculable without 
losing the interest of being genuinely non-perturbative in its origin 
(see~\cite{hw} for a discussion of the new hard scale in a framework close 
to the present paper). It remains to be seen in how far this large-target 
approach will allow for a description of the qualitative features in the 
realistic proton case. 

The paper is organized as follows. In Sect.~\ref{lor} the scattering of 
a scalar photon off the target color field is calculated in the parton 
model and the semiclassical approach. 
The comparison of the two results gives rise to the leading-order 
semiclassical expression for the gluon distribution. 
In Sects.~\ref{scnlo} and~\ref{pmnlo}, the semiclassical and parton model 
calculations, respectively, are carried out at next-to-leading order. The 
extraction of the next-to-leading order contribution to the gluon 
distribution from the comparison of the semiclassical and the parton model 
results is the subject of Sect.~\ref{extg}. Section~\ref{conc} contains our 
conclusions, and a number of technical details of the calculations are 
outlined in Appendices~\ref{nvf}--\ref{simp}.

%%%%%%%%%%%%%%%%%%%%%%%%%%%%%%%%%%%%%%%%%%%%%%%%%%%%%%%%%%%%%%%%%%%%%%%%
\section{The leading-order result}\label{lor}

In the following analysis we use a scalar `photon' coupled directly to 
the gluon field as a convenient theoretical tool for extracting the gluon 
distribution~\cite{mue}. 
To be precise, the real `photon' field $\chi$ couples to the field strength 
tensor $F_{\mu\nu}$ via the interaction Lagrangian 
\be
{\cal L}_I=-\frac{\lambda}{2}\,\chi\,\mbox{tr}F_{\mu\nu}F^{\mu\nu}\,.
\ee

\begin{figure}[ht]
\begin{center}
\vspace*{.2cm}
\parbox[b]{4cm}{\psfig{width=4cm,file=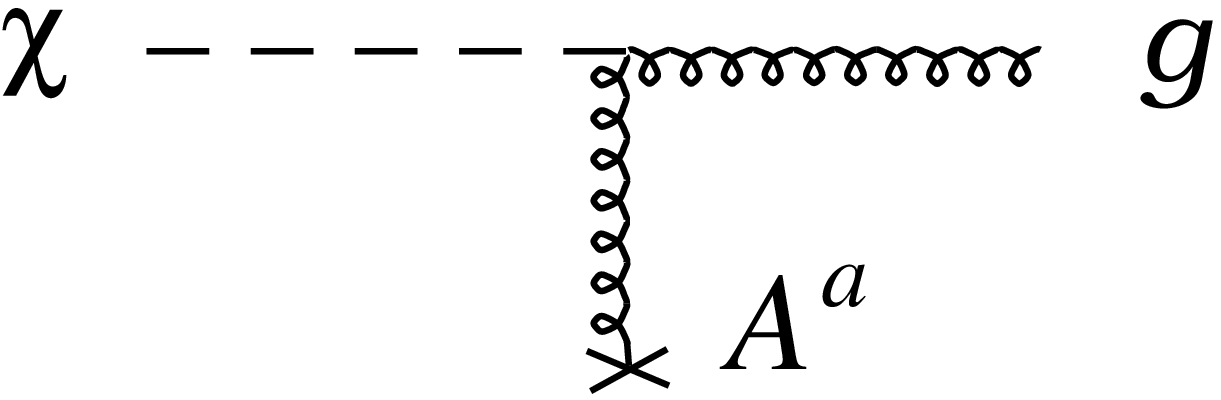}}\\
\end{center}
\refstepcounter{figure}\label{clo}
{\bf Figure \ref{clo}:} The process $\chi\to g$ in an external color field 
with one gluon exchange. 
\end{figure}

The leading-order amplitude for the scattering of the `photon' off a 
classical color field is given by the diagram in Fig.~\ref{clo}. 

Let $q$ and $k$ be the momenta of the incoming virtual `photon' 
($q^2=-Q^2$) and the outgoing gluon respectively. 
We define the light-cone components of a vector $p$ by 
$p_\pm=p_0\pm p_3$ and work in a frame where the plus components of 
$q$ and $k$ are large. In the high-energy limit, the amplitude ${\cal T}^a$ 
corresponding to Fig.~\ref{clo} is given in the rest frame of the proton by 
\be
i\,2\pi\,\delta(k_0-q_0)\,{\cal T}^a(\Delta_\perp)=-i\lambda\,\left(
\frac{1}{2}k_+\tilde{A}_-^a(\Delta)\right)\,(\epsilon_\perp^*\Delta_\perp) 
\,.\label{amp}
\ee
Here $\Delta=k-q$, the field $\tilde{A}$ is the Fourier transform of the 
external color field $A$, and $\epsilon$ is the polarization vector of the 
produced gluon. The evaluation of the r.h. side of Eq.~(\ref{amp}) in the 
high-energy limit shows that ${\cal T}^a$ does not depend on $\Delta_+$ and 
$\Delta_-$ due to the softness of the external field. 
It is convenient to consider the impact parameter space amplitude
\be
\tilde{\cal T}^a(x_\perp)=\frac{i\lambda q_0}{2 C_A}\int dx_+\,\mbox{tr} 
\Big[T^a(\epsilon_\perp^*\partial_\perp)A^{\cal A}_-(x_+,x_\perp)
\Big]\label{tt}  \ee
(with $\partial_\perp\equiv \partial/\partial x_\perp$), which is related to 
the amplitude in Eq.~(\ref{amp}) by a Fourier transformation in transverse 
space. The $x_-$ dependence
of $A^{\cal A}$ is irrelevant in  the high-energy limit.
Here $A^{\cal A}=A^bT^b$ and $T^b$ are the generators of $SU(N_c)$ in the 
adjoint representation, which we use throughout this paper; $C_A = N_c$.  

\begin{figure}[ht]
\begin{center}
\vspace*{.2cm}
\parbox[b]{4cm}{\psfig{width=4cm,file=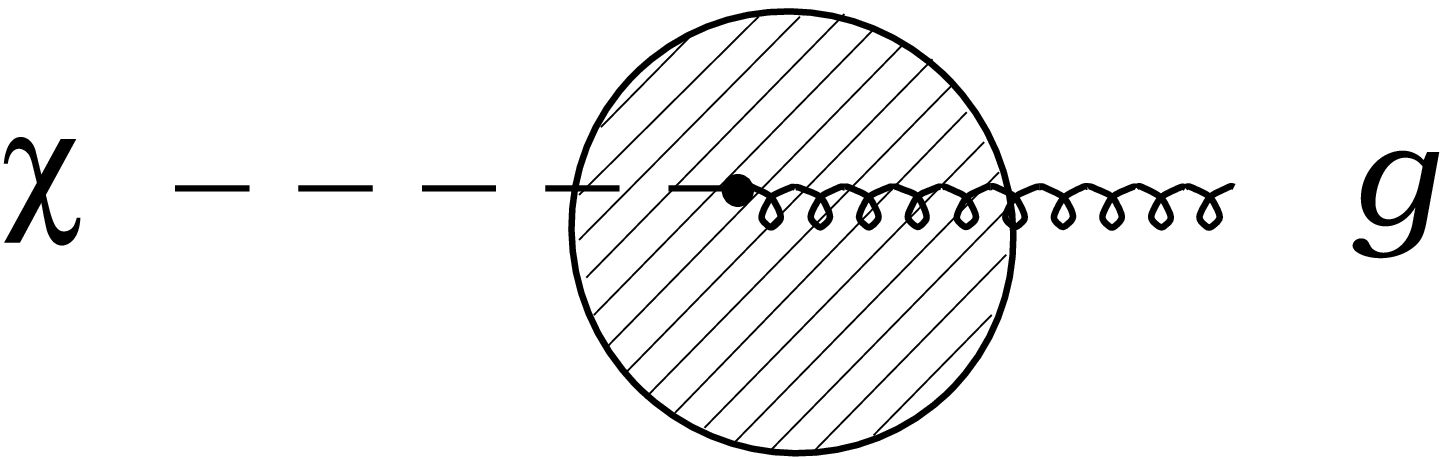}}\\
\end{center}
\refstepcounter{figure}\label{cao}
{\bf Figure \ref{cao}:} The process $\chi\to g$ in an external color 
field. The dot symbolizes the initial $\chi gg$ vertex. 
\end{figure}

Resumming the gluon exchange to all orders means that the fast gluon 
created at the initial $\chi gg$ vertex acquires a non-Abelian eikonal 
factor while travelling through the rest of the external field. 
This is illustrated in Fig.~\ref{cao}. 

The transition to the resummed amplitude of Fig.~\ref{cao} is realized by 
the substitution 
\be
\partial_\perp A^{\cal A}_- \,\,\longrightarrow\,\, U^{\cal A}_{(\infty,x_+)}(x_\perp)\,
\partial_\perp A^{\cal A}_-\,U^{{\cal A} \dagger}_{(\infty,x_+)}(x_\perp)\,
\ee
in Eq.~(\ref{tt}), where 
\be
U^{\cal A}_{(\infty,x_+)}(x_\perp)=P\exp\left[-\frac{ig}{2}\int_{x_+}^\infty 
dx_+A^{\cal A}_-(x_+,x_\perp)\right] .
\label{umatrix}
\ee
The operator $P$ denotes path ordering along $x_+$. 

From this, the total cross section for the scattering of the virtual photon 
off the color field target, i.e., the semiclassical $(sc)$ leading-order 
result, can be derived. Using the identity 
\be
\int_{-\infty}^\infty dx_+\,U^{\cal A}_{(\infty,x_+)}(x_\perp)\,
(\partial_\perp A^{\cal A}_-(x_+,x_\perp))\,U^{\cal A}_{(x_+,-\infty)}
(x_\perp)=\left(-\frac{2}{ig}\right)\,\partial_\perp 
U^{\cal A}_{(\infty,-\infty)}(x_\perp)\,,
\ee
one finds
\be
\sigma^{(0)}_{sc}(x,Q^2)=\frac{\lambda^2}{4g^2 C_A}\int d^2x_\perp\left|
\frac{\partial}{\partial y_\perp}W^{\cal A}_{x_\perp}(y_\perp)\Big|_{y_\perp=0}
\right|^2\equiv\frac{\lambda^2}{4g^2C_A}\int d^2x_\perp|\partial_\perp 
W^{\cal A}_{x_\perp}(0_\perp)|^2\,,\label{sc0}
\ee
where the index (0) stands for `leading order'. Here
\be
W^{\cal A}_{x_\perp}(y_\perp)=
U^{\cal A}(x_\perp)U^{{\cal A} \dagger}(x_\perp+y_\perp)-1\label{wdef} \ee
and
\be
U^{\cal A}(x_\perp)=U^{\cal A}_{(\infty,-\infty)}(x_\perp)\,.
\label{umatrixk}
\ee
In Eq.~(\ref{sc0}), the summation over color indices is implicit. In the 
following it is always assumed that an appropriate averaging over the 
color fields underlying the basic quantity $W^{\cal A}$ is performed 
(cf.~\cite{bgh}). 

\begin{figure}[ht]
\begin{center}
\vspace*{.2cm}
\parbox[b]{3.5cm}{\psfig{width=3.5cm,file=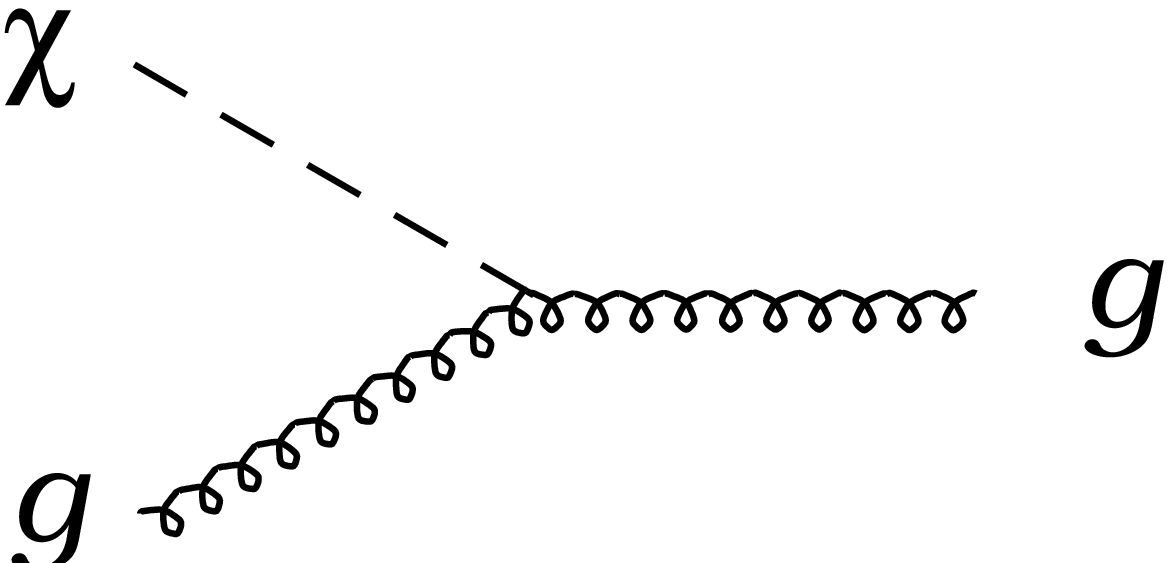}}\\
\end{center}
\refstepcounter{figure}\label{par}
{\bf Figure \ref{par}:} The partonic process $\chi g\to g$ at leading 
order.
\end{figure}

The leading-order semiclassical gluon distribution is obtained by comparing 
this result with a conventional partonic calculation, where the target is 
described by a gluon distribution. The relevant diagram is shown in 
Fig.~\ref{par}, and the corresponding parton model $(pm)$ cross section 
reads 
\be
\sigma^{(0)}_{pm}(x,Q^2)=\frac{\pi\lambda^2}{4}\,xg^{(0)}(x,Q^2)\,.
\label{pm0}
\ee

Identifying the cross sections of Eqs.~(\ref{sc0}) and (\ref{pm0}), one 
obtains 
\be
xg^{(0)}(x,Q^2)=\frac{1}{2\pi^2\alpha_s}\frac{1}{2 C_A} \int
d^2x_\perp|\partial_\perp  W^{\cal A}_{x_\perp}(0_\perp)|^2\,.\label{gdlo}
\ee
This result has been derived in~\cite{bgh} identifying the scaling 
violations of $F_2$ with the gluon distribution. 
Note the color factor $\frac{1}{2 C_A}$ in Eq.~(\ref{gdlo}) due to the 
adjoint representation, which we use throughout the paper. As expected, 
this leading-order gluon distribution $xg^{(0)}(x,Q^2)$ is constant for 
$x\to 0$  and shows no scaling violations beyond those induced by the 
explicit  $\alpha_s$-factor on the r.h. side of Eq.~(\ref{gdlo}). 
In~\cite{dgp} the value of $\lim_{x\to 0} xg(x,Q^2)$ has been given in the 
stochastic vacuum model. 

%%%%%%%%%%%%%%%%%%%%%%%%%%%%%%%%%%%%%%%%%%%%%%%%%%%%%%%%%%%%%%%%%%%%%%%%%
\section{Semiclassical calculation at next-to-leading order}\label{scnlo}

The leading ln$\,Q^2$ calculation of the process $\chi\to gg$ in the 
semiclassical approach reproduces the conventional gluon-gluon splitting 
function, as shown in~\cite{bgh}. Here, we need the complete 
next-to-leading order total cross section for the scattering of a $\chi$ 
particle off an external color field. This calculation can be simplified 
significantly if one starts with a discussion of all diagrams contributing 
to the forward amplitude. The total cross section follows from the 
imaginary part of this amplitude. Furthermore, it is convenient to begin 
by considering the two-gluon exchange approximation. 

\begin{figure}[ht]
\begin{center}
\vspace*{.2cm}
\parbox[b]{13cm}{\psfig{width=13cm,file=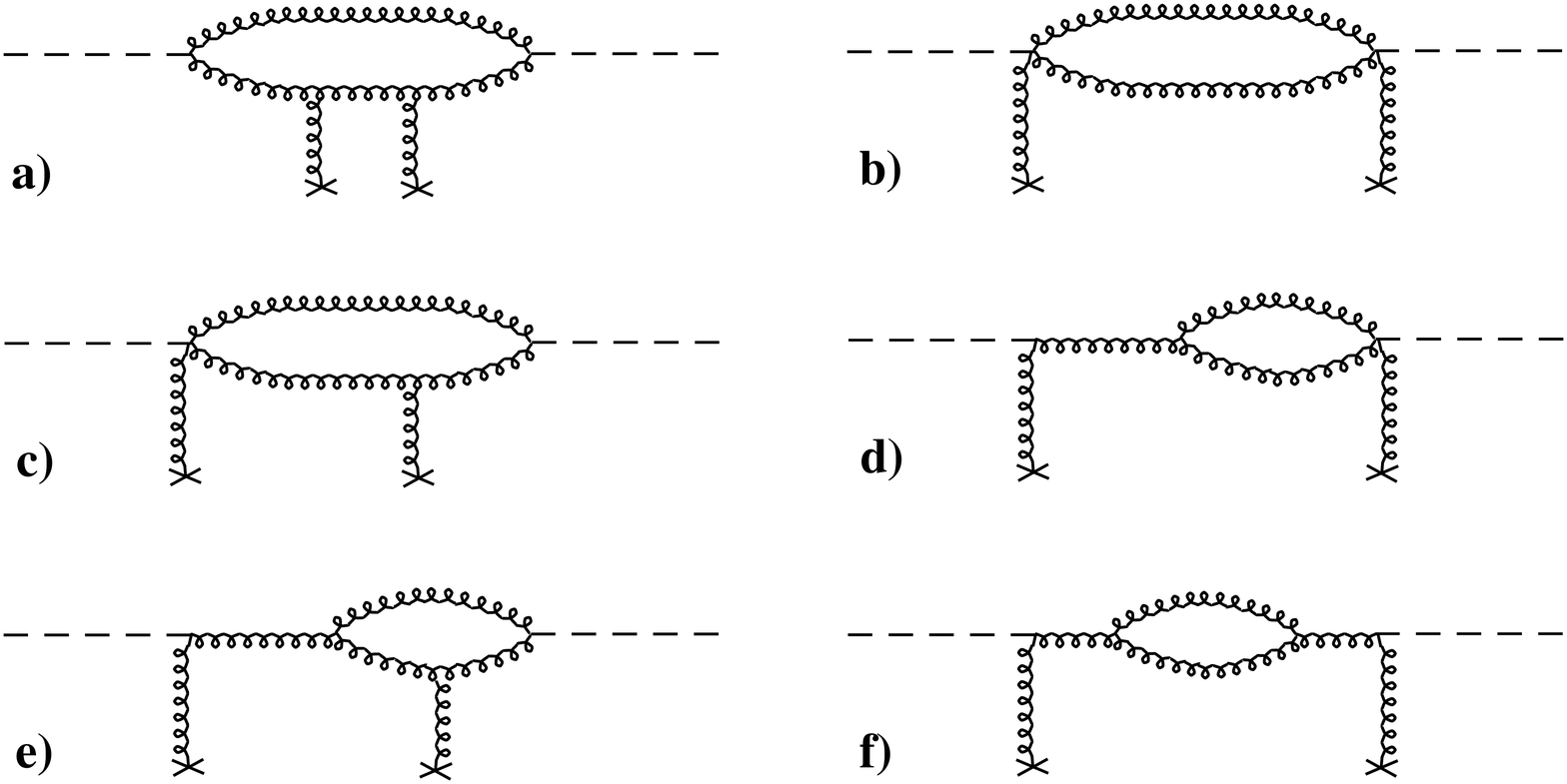}}\\
\end{center}
\refstepcounter{figure}\label{nlo}
{\bf Figure \ref{nlo}:} Two-gluon exchange diagrams for the forward 
scattering of the scalar `photon' $\chi$ off a color field. The mirrored 
analogues of~c),~d) and~e) and the analogue of graph~a) with the two $t$ 
channel gluons attached to different horizontal lines are not shown. 
\end{figure}

Clearly, the leading order diagram for the forward amplitude is simply the 
square of Fig.~\ref{clo}. At next-to-leading order, all the diagrams in 
Fig.~\ref{nlo} have to be considered. 

In the high-energy limit, the diagrams~d),~e) (with their mirrored 
analogues) and diagram~f) do not contribute. This can be understood 
intuitively by recalling that the gluon field is localized in a given 
region of space. Therefore, in the limit of infinite plus momentum, the 
right-moving gluonic degrees of freedom have no time for a virtual 
fluctuation between their first and second interaction with the external 
field (see Appendix~\ref{nvf} for a more technical argument). A related 
discussion in the case of particle radiation in high-energy scattering of 
external fields can be found in~\cite{bh}. 

\begin{figure}[ht]
\begin{center}
\vspace*{.2cm}
\parbox[b]{13cm}{\psfig{width=10cm,file=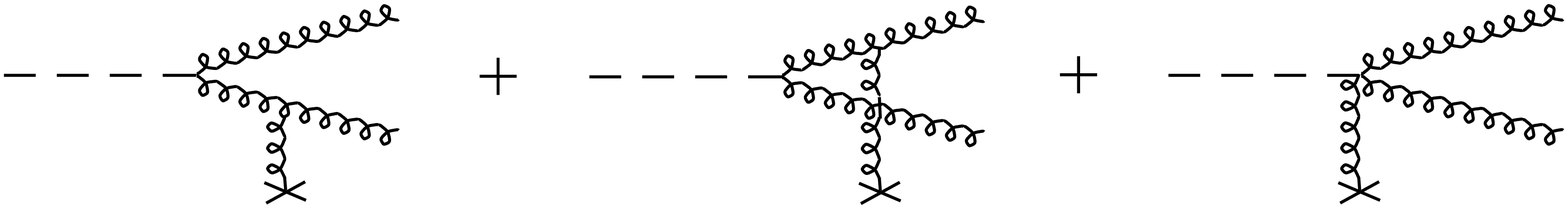}}\\
\end{center}
\refstepcounter{figure}\label{thd}
{\bf Figure \ref{thd}:} The simplified amplitude for $\chi\to gg$, which 
can be used for the calculation of the total cross section.
\end{figure}

What remains to be calculated is the imaginary part of diagram~a) (and the 
corresponding diagram where the two $t$ channel gluons are attached to 
different lines), diagram~b) and diagram~c) (with its mirrored analogue). 
This corresponds to calculating the cross section for $\chi\to gg$ from the 
amplitude defined by the three diagrams in Fig.~\ref{thd}. 
Note that this is a significant simplification since the amplitude in 
Fig.~\ref{thd} has no contributions arising from cutting diagrams~d),~e) 
and~f) of Fig.~\ref{nlo}.

\begin{figure}[ht]
\begin{center}
\vspace*{.2cm}
\parbox[b]{13cm}{\psfig{width=10cm,file=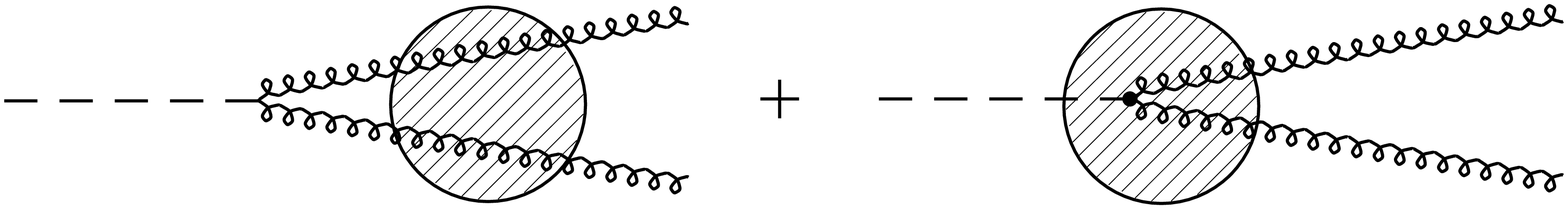}}\\
\end{center}
\refstepcounter{figure}\label{scc}
{\bf Figure \ref{scc}:} The relevant contributions to the amplitude in the 
next-to-leading order semiclassical calculation. The dot in the second 
diagram symbolizes the initial $\chi ggg$ vertex. 
\end{figure}

The next step is to resum the interaction with the external field to all 
orders, i.e., to repeat the step leading from Fig.~\ref{clo} to 
Fig.~\ref{cao}. 
The result is illustrated in Fig.~\ref{scc}. 
Here, the first diagram corresponds to the fluctuation of the incoming 
`photon' into a gg pair before the target and the subsequent scattering 
of the gluons off the color field, treated in the eikonal approximation. 
In an expansion in powers of the external field, the leading term reproduces 
the first two diagrams of Fig.~\ref{thd}. The second diagram of 
Fig.~\ref{scc} corresponds to the creation of the two gluons in the 
space-time region of the external field, via a $\chi ggg$ vertex. The two 
fast gluons then acquire non-Abelian eikonal phases while travelling 
through the rest of the color field. An expansion in powers of the 
external field generates the third diagram of Fig.~\ref{thd}. 

The amplitude of Fig.~\ref{scc} can be calculated using the methods 
of~\cite{bdh} (see also~\cite{h}). 
Some details relevant to this particular process are discussed in 
Appendix~\ref{calc}. 
The result can be given in the form 
\be
\sigma^{(1)}_{sc}(x,Q^2)=\frac{\lambda^2}{32(2\pi)^6}\int \!\!d\alpha\!
\int\!\!\frac{dk_\perp'^2}{\alpha(1-\alpha)}\int \!\!d^2x_\perp\left|\int 
d^2 k_\perp\frac{N^2\delta_{ij}+2k_ik_j}{N^2+k_\perp^2}\,
\tilde{W}^{\cal A}_{x_\perp}(k_\perp'\!-\!k_\perp)\right|^2\!\!,
\label{sc1}
\ee
where $\alpha$ and $1-\alpha$ are the longitudinal momentum fractions 
carried by the two gluons, $N^2=\alpha(1-\alpha)Q^2$, $k_\perp$ and 
$k_\perp'$ are the transverse momenta of one of the two gluons before and 
after the interaction with the external field respectively, and
$\tilde{W}^{\cal A}$  is the Fourier transform of the function defined in
Eq.~(\ref{wdef}). % The 
%index ${\cal A}$ signifies that the two matrices $U^{\cal A}$ and $U^{{\cal A} \dagger}$
%have  to be taken in the adjoint representation. 
Summation over the indices
$i,j$  and over the color indices of $\tilde{W}^{\cal A}$ is implicit. 

The full next-to-leading order semiclassical result is given by the sum 
of Eqs.~(\ref{sc0}) and~(\ref{sc1}):
\be 
\sigma_{sc}(x,Q^2)=\sigma^{(0)}_{sc}(x,Q^2)+\sigma^{(1)}_{sc}(x,Q^2)\,.
\label{scs}
\ee

%%%%%%%%%%%%%%%%%%%%%%%%%%%%%%%%%%%%%%%%%%%%%%%%%%%%%%%%%%%%%%%%%%%%%%%%%%
\section{Parton model result at next-to-leading order}\label{pmnlo}

\begin{figure}[ht]
\begin{center}
\vspace*{.2cm}
\parbox[b]{10.8cm}{\psfig{width=10.8cm,file=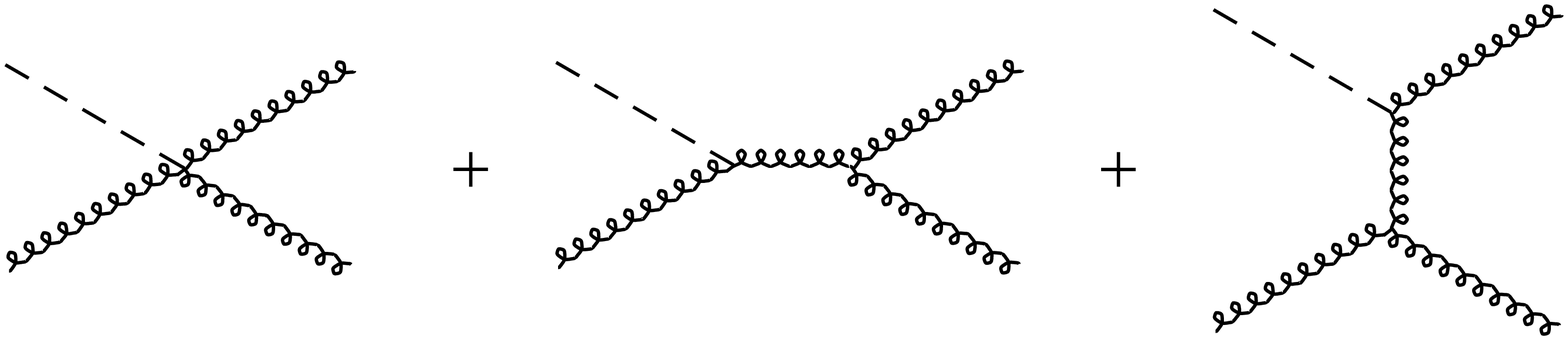}}\\
\end{center}
\refstepcounter{figure}\label{pnlo}
{\bf Figure \ref{pnlo}:} Conventional partonic amplitude for the process 
$\chi g\to gg$. The crossed contribution belonging to the last of the 
three graphs and the ghost diagrams that have to be added for covariant 
polarization summation are not shown. 
\end{figure}

Working in $d=4+\epsilon$ dimensions, the amplitude corresponding to 
Fig.~\ref{pnlo}, which includes a factor $1/(2+\epsilon)$ for initial state 
gluon polarization, a factor $1/2$ for identical final state particles, and 
the color factor $C_A=N_c$, reads 
\be
|{\cal T}_{\chi g\to gg}|^2\,=\,\frac{C_A}{2}\,\lambda_d^2g_d^2\,\left\{
\frac{1}{2\hat{s}\hat{t}\hat{u}}\left(\hat{s}^4+\hat{t}^4+\hat{u}^4+Q^8 
\right)-\frac{2\epsilon}{2+\epsilon}\,Q^2\right\}\,.
\ee
Here $\lambda_d=\lambda\mu^{-\epsilon/2}$ and $g_d=g\mu^{-\epsilon/2}$ are 
the $d$-dimensional couplings, and $\hat{s}$, $\hat{t}$, $\hat{u}$ are the 
usual Mandelstam variables of a 2-to-2 scattering process. 

{}From this squared amplitude, the total partonic cross section 
$\hat{\sigma}^{(1)}(z,Q^2)$ for the process $\chi g\to gg$ is obtained by 
standard methods. The variable $z$ is defined by $z=Q^2/(Q^2+\hat{s})$. 
Alternatively, it is given by $z=x/y$, where $y$ is the fraction of the 
target momentum carried by the struck gluon. Combining this with the 
leading order result, one can write 
\be
\sigma_{pm}(x,Q^2)=\int_x^1\frac{dz}{z}\left(\sigma_{0\,\{d\}}\,
\delta(1-z)+\hat{\sigma}^{(1)}(z,Q^2)\right)\,yg_b(y)\,,
\ee
where $\sigma_{0\,\{d\}}=\pi\lambda_d^2/4$ (cf.~Eq.~(\ref{pm0})). Note that 
in the following we only calculate terms enhanced by powers of $\ln(1/x)$. 
Therefore virtual corrections, which affect only the endpoint $z=1$ and 
do not produce such terms, do not contribute. 

Working in the $\msb$ scheme, we renormalize the bare gluon distribution 
$g_b(x)$ according to
\be
g_b(x)=g(x,\mu^2)-\frac{\alpha_s}{2\pi}\int_{x}^{1}\frac{dz}{z}P_{gg}(z)
\left\{\frac{2}{\epsilon}+\gamma_E-\ln 4\pi\right\}\,g(y,\mu^2) \,.
\ee
The result can finally be given in the form
\be
\sigma_{pm}(x,Q^2)=\sigma_{0\,\{d=4\}}\,x\,\int_x^1\frac{dz}{z}\left(\delta
(1-z)+\frac{\alpha_s}{2\pi}\bigg[P_{gg}(z)\ln\frac{Q^2}{\mu^2}+
C^{\mmsb}_g(z)\bigg]\right)g(y,\mu^2)\,,\label{pms}
\ee
where the integrand is only complete in the region $z<1$. In 
Eq.~(\ref{pms}) $P_{gg}(z)$ (with $z<1$) is the usual gluon-gluon splitting 
function. The coefficient function $C^{\mmsb}_g(z)$, characteristic of the 
process under consideration, has been derived to be 
\be
C^{\mmsb}_g(z)\,=\,P_{gg}(z)\ln\frac{1-z}{z}\,-\,\frac{11C_A}{6z(1-z)}\,.
\ee

%%%%%%%%%%%%%%%%%%%%%%%%%%%%%%%%%%%%%%%%%%%%%%%%%%%%%%%%%%%%%%%%%%%%%%%%%%%
\section{Extracting the gluon distribution}
\label{extg}

As in the leading-order case discussed in Sect.~\ref{lor}, the gluon 
distribution is extracted from the next-to-leading order semiclassical 
calculation by identifying the next-to-leading order total cross sections 
of the semiclassical and the parton model approach given by 
Eqs.~(\ref{scs}) and (\ref{pms}): $\sigma_{sc}(x,Q^2)=\sigma_{pm}(x,Q^2)$. 

We write the gluon distribution $xg(x,\mu^2)$ entering Eq.~(\ref{pms}) as 
\be 
xg(x,\mu^2)=xg^{(0)}(x,\mu^2)+xg^{(1)}(x,\mu^2)\,,\label{gds}
\ee
where $xg^{(0)}(x,\mu^2)$ is given by Eq.~(\ref{gdlo}) and $xg^{(1)} 
(x,\mu^2)$ is a higher-order correction. 
Inserting Eq.~(\ref{gds}) into Eq.~(\ref{pms}), both $g^{(0)}$ and 
$g^{(1)}$ are kept in the $\delta(1-z)$ term, but only the leading-order 
distribution $g^{(0)}$ is kept in the $\alpha_s$ contribution. 
Now, identifying $\sigma_{sc}$ and $\sigma_{pm}$, the leading order parts 
of both cross sections cancel and one finds 
\bea
xg^{(1)}(x,\mu^2)&=&\frac{1}{4(2\pi)^7}\int \!\!d\alpha\!
\int\!\!\frac{dk_\perp'^2}{\alpha(1-\alpha)}\int \!\!d^2x_\perp\left|\int 
d^2 k_\perp\frac{N^2\delta_{ij}+2k_ik_j}{N^2+k_\perp^2}\,
\tilde{W}^{\cal A}_{x_\perp}(k_\perp'\!-\!k_\perp)\right|^2\nonumber
\\
&&-\frac{\alpha_s}{2\pi}\int_x^1dz\left[P_{gg}(z)\ln\frac{Q^2}
{\mu^2}+C^{\mmsb}_g(z)\right]yg^{(0)}(y,\mu^2)\,.\label{g1}
\eea
When evaluating the r.h. side of this equation, all terms that are not 
enhanced by $\ln(1/x)$ will be dropped. 
In particular, we can use
\be
P_{gg}(z)\simeq\frac{2C_A}{z}\qquad\mbox{and}\qquad C_g^{\mmsb}(z)\simeq
\frac{2C_A}{z}\left(\ln\frac{1}{z}-\frac{11}{12}\right)\,.
\ee
Note that in Eq.~(\ref{g1}) we are only interested in the leading twist 
contribution of the semiclassical cross section which will be determined 
in Eq.~(\ref{wint}).

The integral involving the function $\tilde{W}^{\cal A}$ is conveniently
rewritten  using the integration variable $z=Q^2/(Q^2+M^2)$, where
$M^2=\hat{s}=  k_\perp'^2/ (\alpha(1-\alpha))$ is the invariant mass of the
two gluons in  the final state. We introduce a parameter $\mu'$ such that, for
a soft  hadronic scale $\Lambda$ governing the behaviour of $\tilde{W}^{\cal
A}$, one has  $\Lambda^2{\scriptstyle\ll}\mu'^2\mst{<}{\sim}Q^2$. This allows
us to  decompose the integral according to (see~Appendix~\ref{int} for
details)  \bea
&&\hspace*{-1.5cm}\frac{1}{4(2\pi)^7}\int \!\!d\alpha\!
\int\!\!\frac{dk_\perp'^2}{\alpha(1-\alpha)}\int \!\!d^2x_\perp\left|\int 
d^2 k_\perp\frac{N^2\delta_{ij}+2k_ik_j}{N^2+k_\perp^2}\,
\tilde{W}^{\cal A}_{x_\perp}(k_\perp'\!-\!k_\perp)\right|^2
\nonumber\\\label{wint}
\\
&=&\frac{1}{4\pi^3}\int^1\frac{dz}{z}\ln\frac{Q^2}{z\mu'^2}\,\int d^2 
x_\perp|\partial_\perp W^{\cal A}_{x_\perp}(0_\perp)|^2\,+\,\frac{2}{\pi}
\int^1\frac{dz}{z}\int_0^{\mu'^2}dk_\perp'^2f(k_\perp'^2)\,,\nonumber
\eea
where the non-perturbative $W^{\cal A}$ dependence is encoded in the function
$f$:  \be
\!\!\!\!\!f(k_\perp'^2)=\!\int\!\!\frac{d^2y_\perp}{(2\pi)^2y_\perp^2}\!
\int\!\!\frac{d^2z_\perp}{(2\pi)^2z_\perp^2}\!\int\!\!d^2x_\perp\mbox{tr}
\left[W_{x_\perp}^{\cal A}(y_\perp)W_{x_\perp}^{\cal A\dagger}(z_\perp)
\right]\,e^{ik_\perp'(y_\perp-z_\perp)}\left(\frac{2(y_\perp z_\perp)^2}
{y_\perp^2z_\perp^2}-\!1\!\right) .\label{fdef}
\ee
Here and in the following we disregard all terms suppressed by powers of 
$\Lambda^2/\mu'^2$. The motivation for writing the integral in the form 
given in Eq.~(\ref{wint}) is the explicit separation of the ln$\,Q^2$ term, 
which is multiplied by the short-distance specific function $\partial_\perp 
W^{\cal A}_{x_\perp}(0_\perp)$.

Note that we have not specified the lower bound of the $z$ integrations in 
Eq.~(\ref{wint}). Clearly, the kinematical limit is $z=x$ since the 
invariant mass of the produced $gg$ pair can not be larger then the total 
center-of-mass energy available. However, no such bound appears explicitly 
in the semiclassical treatment since the classical color field behaves 
like an infinitely heavy target. The physical cutoff is provided by the 
breakdown of the semiclassical approximation for $z\sim x$, i.e., $y\sim 1$. 

Therefore, in order to obtain the leading logarithm, the second term on 
the r.h. side of Eq.~(\ref{wint}) has to be treated by applying the 
substitution
\be
\int^1\frac{dz}{z}\,\,\longrightarrow\,\,\ln\frac{1}{x}\,.\label{ls}
\ee
For the first term on the r.h. side of Eq.~(\ref{wint}), only the short 
distance structure of the external field matters. 
According to Eq.~(\ref{gdlo}), this short distance behaviour is 
characterized by the leading order gluon distribution $yg^{(0)}(y,\mu^2)$. 
Thus, we substitute 
\be
\frac{1}{2 C_A} \int^1\frac{dz}{z}\ln\frac{Q^2}{z\mu'^2}\,\int
d^2x_\perp|\partial_\perp 
W^{\cal A}_{x_\perp}(0_\perp)|^2\,\,\longrightarrow\,\,\int_x^1\frac{dz}{z}\ln
\frac{Q^2}{z\mu'^2}\,2\pi^2\alpha_s(\mu^2)\,yg^{(0)}(y,\mu^2)\label{l2s} \ee
and assume that the behaviour of the phenomenological gluon distribution 
at $z\to x$, i.e., $y=x/z\to 1$, correctly accounts for the region where 
the semiclassical treatment is no longer valid. 
In fact, when inserting Eq.~(\ref{wint}) with the substitutions 
Eq.~(\ref{ls}) and~(\ref{l2s}) into Eq.~(\ref{g1}), the details of the 
large-$y$ behaviour of $yg^{(0)} (y,\mu^2)$ do not matter since the 
$\ln^2(1/x)$ enhanced contributions from the semiclassical and the 
partonic calculations cancel. 
Thus, we obtain 
\be
xg^{(1)}(x,\mu^2)=\ln\frac{1}{x}\left[\frac{\alpha_s}{\pi}\,C_A\left(\ln
\frac{\mu^2}{\mu'^2}+\frac{11}{12}\right)xg^{(0)}(x,\mu^2)+\frac{2}{\pi}
\int_0^{\mu'^2}dk_\perp'^2f(k_\perp'^2)\right] ,
\ee
where the $\mu'^2$ dependence cancels between the two terms. 
Therefore, we can set $\mu'^2=\mu^2\exp[11/12]$ and write 
\be 
xg^{(1)}(x,\mu^2)=\frac{2}{\pi}\left(\ln\frac{1}{x}\right)
\int_0^{\mu^2\exp[11/12]}dk_\perp'^2f(k_\perp'^2)\,.\label{g1f}
\ee

Note that the correction $xg^{(1)}(x,\mu^2)$ shows scaling violations 
consistent with the Altarelli-Parisi evolution at small $x$ and a 
logarithmic small-$x$ enhancement that is sensitive to the non-perturbative, 
large-size structure of the target. 

It would be interesting to evaluate Eq.~(\ref{g1f}) in the framework of 
the model of the stochastic vacuum of~\cite{dk} or of the large hadron 
model employed in~\cite{bgh}. 
Note in particular that, following~\cite{bgh}, the large-$N_c$ expression 
\be
\int d^2x_{\perp} \, \mbox{tr}\left(W^{\cal A}_{x_\perp}(y_\perp)
W^{\cal A\dagger}_{x_\perp}(z_\perp)\right)=\Omega N_c^2\left[1-
e^{-ay_\perp^2}-e^{-az_\perp^2}+e^{-a(y_\perp-z_\perp)^2}\right]
\ee
can be used in Eq.~(\ref{fdef}). 
Here $\Omega$ is the geometrical cross section of the target, the 
impact parameter dependence of the target thickness (which would be 
reflected in an impact parameter dependence of the parameter $a$) is 
neglected, and averaging over all relevant color field configurations 
is assumed. 
With this model, the unintegrated, i.e., $k_\perp'$ dependent, version 
of $xg^{(1)}(x,\mu^2)$ reproduces the recent result of Mueller 
(cf.~Eq.~(44) of~\cite{mue1}). 
The result presented in Eq.~(\ref{g1f}) extends the discussion given 
in~\cite{mue1} by carefully matching the semiclassical and the parton 
model calculations. 
This gives rise to a precise definition of the cutoff of the $k_\perp'$ 
integration in Eq.~(\ref{g1f}) and to the interpretation of the result 
as a correction to the leading order gluon distribution of Eq.~(\ref{gdlo}). 

Equations~(\ref{fdef}) and (\ref{g1f}) can be significantly simplified 
(cf. Appendix~\ref{simp} for details) leading to the main result of our 
paper: 
\be
xg^{(1)}(x,\mu^2)=\frac{1}{\pi^3}\left(\ln\frac{1}{x}\right)\,
\int_{r(\mu)^2}^{\infty}\,\frac{dy_\perp^2}{y_\perp^4}\left\{-\int 
d^2x_\perp \,\mbox{tr} W^{\cal A}_{x_\perp}(y_\perp)\right\}\,,\quad r(\mu)^2=
\frac{4e^{\frac{1}{12}-2\gamma_E}}{\mu^2}\,.\label{g1ff}
\ee
It is easy to see how such a formula comes about: if the cutoff $\mu^2$ 
in Eq.~(\ref{g1f}) could be taken to infinity, the integration 
$dk_\perp'^2=d^2k_\perp'/\pi$ would give rise to the $\delta$-function 
$\delta^{2}(y_\perp\!-\!z_\perp)$ (cf. Eq.~(\ref{fdef})), and either the 
$y_\perp$ or the $z_\perp$ integration could be trivially performed. The 
remaining integration, say the $y_\perp$ integration, is now divergent at 
small $y_\perp$, showing that the cutoff in Eq.~(\ref{g1f}) can, in fact, 
not be removed. The integrand of this divergent $y_\perp$ integration is 
precisely the one of Eq.~(\ref{g1ff}) and, as can be seen by closer 
inspection of the relevant integrals (cf. Appendix~\ref{simp}), it is 
possible to translate the upper cutoff of the $k_\perp'$ integration into a 
lower cutoff of the $y_\perp$ integration. This is the origin of the 
compact formula in Eq.~(\ref{g1ff}). 

Given a specific model for the gluon fields of the target that allows for 
the calculation of the fundamental quantity $W^{\cal A}_{x_\perp}(y_\perp)$, 
Eq.~(\ref{g1ff}) can be used to improve the leading order semiclassical 
result of Eq.~(\ref{gdlo}). 

With the large-$N_c$ expression of~\cite{bgh}, 
\be
-2\int d^2x_{\perp} \, \mbox{tr}\,W^{\cal A}_{x_\perp}(y_\perp)
=\int d^2x_{\perp} \, \mbox{tr}\left(W^{\cal A}_{x_\perp}(y_\perp)
W^{\cal A\dagger}_{x_\perp} (y_\perp)\right)
=2\Omega N_c^2\left[1-e^{-ay_\perp^2}\right]\,,
\ee
one can see the qualitative agreement of Eq.~(\ref{g1ff}) with 
Eqs.~(49)--(51) of~\cite{mue}. 
The progress of the present 
investigation is the careful matching of the semiclassical and the 
parton model treatment, giving rise to the precise definition of the cutoff 
of the $y_\perp$ integration in terms of the scale of the gluon 
distribution. 

%%%%%%%%%%%%%%%%%%%%%%%%%%%%%%%%%%%%%%%%%%%%%%%%%%%%%%%%%%%%%%%%%%%%%%%%%
\section{Summary and conclusions}\label{conc}

The semiclassical approach has been successfully applied to different 
kinds of high-energy reactions. In DIS, the interaction of the energetic 
partons of the photon with the target can be calculated in the eikonal 
approximation and is essentially given by a (non-perturbative) 
Wegner-Wilson loop, 
which measures an integral of the field strength of the target. 

This approach, though limited to low scales $\mu^2$, can predict input 
distributions for the DGLAP equation. Of course, in order to make numerical 
estimates of the parton densities, the evaluation of the Wegner-Wilson loop 
in a specific non-perturbative model is necessary.

In this respect the gluon distribution is of particular interest as it 
dominates DIS at low $x$. To leading order, the distribution $x\,g^{(0)} 
(x,\mu^2)$ is a constant characterizing the averaged local field strength 
in the target. Therefore, a calculation at next-to-leading order is 
mandatory to obtain a nontrivial energy dependence of the gluon density. 

Using a scalar `photon' (denoted by $\chi$), which couples directly to the
gluon in a gauge invariant way, we derive the gluon density by matching 
the semiclassical and the parton model approach. At leading order, we have 
to equate the cross section for the transition $\chi \to g$ in a soft 
external field with the cross section of the process $\chi g \to g$ as 
given in the parton model. At next-to-leading order, the cross section for 
the transition $\chi \to gg$ in a soft external field has to be equated 
with the parton model cross section of the process $\chi g \to gg$. The 
$\alpha_s$ correction of the gluon distribution shows a logarithmic 
enhancement, i.e., $x\,g^{(1)}(x,\mu^2) \propto \ln(1/x)$. 

In our approach we can not go beyond the leading-ln$(1/x)$ approximation. 
This is a fundamental limitation of the semiclassical treatment and many 
related approaches, which has its origin in the artificial separation of 
the QCD dynamics into the soft degrees of freedom of the target and the 
high-energy modes of the projectile. Integrating over the fluctuations of 
the projectile, one has to drop the soft modes since their interaction can 
not be treated in the eikonal approximation. 

Parton distributions at next-to-leading order are scheme dependent. Our 
final result in Eq.~(\ref{g1ff}) (together with the leading order 
contribution of Eq.~(\ref{gdlo})) provides the gluon density in the $\msb$ 
scheme. The short-distance cutoff in Eq.~(\ref{g1ff}) is quantitatively 
related to the scale of the gluon distribution. This enables us to obtain 
numerical predictions for the gluon distribution at next-to-leading order 
in any non-perturbative approach describing the soft color field of the 
proton.\\[.4cm]
{\bf Acknowledgements}\\[.1cm]
One of us (A.H.) would like to thank W. Buchm\"uller and T. Gehrmann for 
the fruitful collaboration on a previous paper, during which essential 
ideas underlying this present investigation took shape. 

%%%%%%%%%%%%%%%%%%%%%%%%%%%%%%%%%%%%%%%%%%%%%%%%%%%%%%%%%%%%%%%%%%%%%%%%%
\begin{appendix}
\section{Suppression of intermediate virtual fluctuations in the 
high-energy limit}\label{nvf}

\begin{figure}[ht]
\begin{center}
\vspace*{.2cm}
\parbox[b]{6cm}{\psfig{width=6cm,file=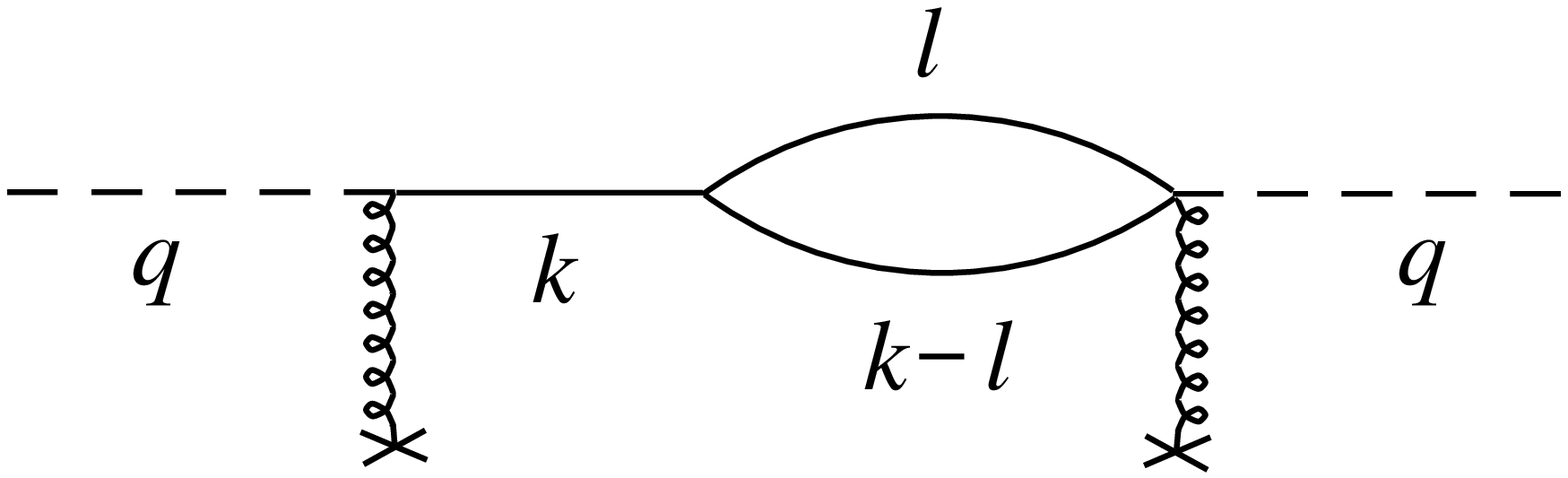}}\\
\end{center}
\refstepcounter{figure}\label{sca}
{\bf Figure \ref{sca}:} Scalar version of diagram Fig.~\ref{nlo}d). 
\end{figure}

To illustrate the vanishing of diagrams d), e) and f) of Fig.~\ref{nlo} in 
the high-energy limit, consider first the simpler case where the horizontal 
gluon lines are replaced by scalar lines. 
The scalar version of diagram Fig.~\ref{nlo}d) with the naming of 
momenta used in this appendix is shown in Fig.~\ref{sca}. 

In the high-energy limit, the soft external field can not change the 
momentum plus component of the fast right-moving particles essentially, 
$k_+\simeq q_+$. 
Furthermore, the dependence of the external field vertex on the minus 
component of the transferred momentum can be neglected. 
Thus, the minus component integrations in the diagram Fig.~\ref{sca} take 
the form
\be
I\!=\!\int\! dk_-\!\int\! dl_-\,\frac{1}{[k_+k_--k_\perp^2+i\epsilon]\,
\,[l_+l_--l_\perp^2+i\epsilon]\,\,[(k\!-\!l)_+(k\!-\!l)_--(k\!-\!l)_\perp^2
+i\epsilon]}\,.
\ee
Note first that this expression vanishes unless $0\le l_+\le k_+$, in which 
case the two poles in the complex $l_-$ plane lie on different sides of the 
integration contour. However, this means that both $k_+$ and $(k-l)_+$ have 
to be positive, so that now both poles in the complex $k_-$ plane are on 
the same side of the integration contour. This demonstrates that the 
integral $I$ vanishes in any case. 

The same argument applies to the scalar analogues of diagrams 
Fig.~\ref{nlo}e) and f), since there the momentum $k$ also flows through 
more than one propagator between the two external field vertices. 
This results in the presence of several poles in $k_-$, all on the same 
side of the integration contour, and therefore in the vanishing of the 
integral. 

All that has been said above applies to the gluonic case as well, as long 
as there is no additional minus component dependence introduced by the 
numerator factors. Such a dependence would prevent us from closing the 
integration contour. Working in Feynman gauge and applying the 
decomposition of the gluon propagator given in Appendix~\ref{calc} to the 
lines with momentum $k$, one can easily convince oneself that, in the 
high-energy limit, there is indeed no dependence on the minus 
components. 

Thus, we are justified in disregarding diagrams Fig.~\ref{nlo}d) -- f) 
when calculating the forward scattering amplitude at next-to-leading 
order.

%%%%%%%%%%%%%%%%%%%%%%%%%%%%%%%%%%%%%%%%%%%%%%%%%%%%%%%%%%%%%%%%%%%%%%%%
\section{Details of the next-to-leading order semiclassical calculation}
\label{calc} 
The l.h. diagram of Fig.~\ref{scc} contains three contributions: two of 
them describe the interaction of a single outgoing gluon with the color
field, whereas in the third case both gluons interact with the field. 
Here we consider the latter case as an example, and outline some crucial 
steps of the evaluation of the corresponding amplitude which we call 
${\cal T}^{gg}$.

Working in Feynman gauge and denoting the momenta of the gluons 
before (after) the interaction by $k$ and $p$ ($k'$ and $p'$), 
respectively, the amplitude reads
\be
i 2\pi \delta(k'_0+p'_0-q_0) {\cal T}_{ij}^{gg} = 
\int\frac{d^4k}{(2\pi)^4} \epsilon_{(i)}^{\alpha\ast}(p') 
V_{\alpha\mu}(p',p) \frac{-i g^{\mu\rho}}{p^2} H_{\rho\sigma}(p,k) 
\frac{-i g^{\sigma\nu}}{k^2} V_{\nu\beta}^{\dagger}(k',k) 
\epsilon_{(j)}^{\beta\ast}(k')\,,\label{amplitude}
\ee
where the indices $i,j$ characterize the polarization of the gluons 
in the final state. The quantity $H$ represents the $\chi gg$ vertex, 
\be
H_{\rho\sigma}(p,k) = i\lambda((kp) \, g_{\rho\sigma}  
 - k_{\rho} p_{\sigma}) \,,
\ee
while $V(p',p)$ and $V(k',k)$, which will be explicitly given below, are 
the effective vertices for the scattering of the fast gluons off the 
external field. 

To simplify the calculation we exploit an alternative representation
of the metric tensor appearing in the gluon propagators. 
For instance, the tensor $g^{\mu\rho}$ of the propagator 
$-i g^{\mu\rho}/p^2$ is decomposed according to
\be
g^{\mu\rho} = -\sum_{i=1}^{2} \epsilon_{(i)}^{\mu}(p)
\epsilon_{(j)}^{\rho}(p) +\frac{n^{\mu}p^{\rho}}{np} 
+ \frac{p^{\mu}n^{\rho}}{np} - \frac{n^{\mu}n^{\rho}}{(np)^2}p^2 \,,
\label{decomp}
\ee
with the light like vector $n=(n_+,n_-,n_{\perp})=(0,2,0_{\perp})$. A 
possible choice of the polarization vectors $\epsilon_{(i)}(p)$ in 
Eq.~(\ref{decomp}), which in particular satisfy $\epsilon_{(i)}(p) \, 
p = 0$, is given by
\be
\epsilon_{(i)}(p) = \biggl(0, \frac{2\,p_{\perp} \epsilon_{(i)\perp}}{p_+}, 
\epsilon_{(i)\perp} \biggr) ,
\ee
with the transverse basis $\epsilon_{(1)\perp} = (1,0)$ and
$\epsilon_{(2)\perp} = (0,1)$.

In the high energy limit, the $t$ channel exchange of gluons leads to
an amplitude which is proportional to $q_+$.
As a consequence, in order to obtain the leading part of the amplitude,
in Eq.~(\ref{decomp}) only the term containing the polarization vectors
has to be considered.
Because of gauge invariance of the $\chi gg$ vertex the second term 
on the r.h. side of Eq.~(\ref{decomp}) vanishes.
The contribution of the remaining two terms is at most proportional to 
$(q_+)^0$ and hence negligible.

The effective vertices are now multiplied by the polarization vectors of 
two on-shell gluons;
these vertices are governed by the non-Abelian eikonal factor 
defined in Eqs.~(\ref{umatrixk}) and (\ref{umatrix}).
One finds
\be
\epsilon_{(i)}^{\alpha}(p') V_{\alpha\mu}(p',p) \epsilon_{(j)}^{\mu}(p) 
= 2\pi \delta(p_0'-p_0) 2p_0 \Bigl[ \tilde{U}^{\cal A}(p_\perp'-p_\perp)
-(2\pi)^2 \delta^2(p_\perp'-p_\perp) \Bigr] \delta_{ij} \,,
\label{vertg}
\ee
and an analogous expression for the second vertex.
The contribution proportional to $\delta^2(p_\perp'-p_\perp)$ subtracts
the term in $\tilde{U}^{\cal A}$ which contains no interaction,
while $\delta_{ij}$ indicates conservation of the gluon helicity.
In  Ref.~\cite{h} the interested reader can find more details of the 
derivation of Eq.~(\ref{vertg}).

Writing the loop integration in terms of light cone variables,
$d^4k = (1/2)dk_+ dk_- d^2k_\perp$, and using the approximation
$\delta(k_0'-k_0) \simeq 2\delta(k_+'-k_+)$ we arrive at
\bea
{\cal T}_{ij}^{gg} & = & -\frac{\lambda}{2(2\pi)^3} q_+^2 
\int dk_- d^2k_\perp \frac{1}{k^2p^2} 
\Bigl[ \alpha(1-\alpha)(Q^2+k^2+p^2)\delta_{ij} + 2k_ik_j \Bigr]
\label{ampa} \\
& &\quad \times \Bigl[ \tilde{U}^{\cal A}(p_\perp'-p_\perp) - 
(2\pi)^2 \delta^2(p_\perp'-p_\perp) \Bigr]
\Bigl[ \tilde{U}^{{\cal A} \dagger}(k_\perp'-k_\perp)
-(2\pi)^2 \delta^2(k_\perp'-k_\perp) \Bigr] \,.
\nonumber 
\eea
The diagram on the r.h. side of Fig.~\ref{scc} cancels the $k^2$ and 
$p^2$ term in the expression $(Q^2+k^2+p^2)$ of Eq.~(\ref{ampa}). Now, 
the $k_-$ integration can be performed by closing the integration contour 
in the lower half of the complex $k_-$ plane. 

The resulting expression contains terms proportional to 
$\tilde{U}^{\cal A} \tilde{U}^{{\cal A} \dagger}$, $\tilde{U}^{\cal A}$, 
and $\tilde{U}^{{\cal A} \dagger}$, as well as a constant term. The two 
additional contributions, where only a single gluon interacts with the 
external field and which belong to the l.h. side of Fig.~\ref{scc}, 
contain terms proportional to $\tilde{U}^{\cal A}$ and 
$\tilde{U}^{{\cal A} \dagger}$ and a constant term. If one adds these 
contributions, the color field dependence of the total amplitude 
${\cal T}_{ij}$ turns out to be 
\bea
&& \tilde{U}^{\cal A}(p_\perp'-p_\perp) 
\tilde{U}^{{\cal A} \dagger}(k_\perp'-k_\perp)
- (2\pi)^4 \delta^2(p_\perp'-p_\perp) \delta^2(k_\perp'-k_\perp) 
\vphantom{\frac{1}{1}}\\
&& \qquad = \int d^2 x_{\perp} e^{-ix_{\perp}(k_{\perp}'+p_{\perp}')}
\tilde{W}_{x_\perp}^{\cal A}(k_{\perp}'-k_{\perp}) \,. \nonumber
\eea
This provides us with the final result of the total amplitude,
\be
{\cal T}_{ij} = \frac{-i\lambda q_+}{2(2\pi)^2} \int d^2 k_{\perp} 
\frac{N^2 \delta_{ij} + 2k_ik_j}{N^2+k_{\perp}^2} 
\int d^2 x_{\perp} e^{-ix_{\perp}(k_{\perp}'+p_{\perp}')}
\tilde{W}_{x_\perp}^{\cal A}(k_{\perp}'-k_{\perp}) \,.
\ee

The cross section in Eq.~(\ref{sc1}) is obtained by means of the 
standard formula for scattering off an external field,
\be
\sigma_{sc}^{(1)}(x,Q^2) = \frac{1}{2} \int 
\frac{d^3 k'}{(2\pi)^3 2k_0'} \frac{d^3 p'}{(2\pi)^3 2p_0'} \,
\frac{1}{2q_0} \, 2\pi \delta(k_0'+p_0'-q_0) |{\cal T}_{ij}|^2 \,,
\ee
where the two identical particles in the final state require the factor 
$1/2$ in front of the integral.

%%%%%%%%%%%%%%%%%%%%%%%%%%%%%%%%%%%%%%%%%%%%%%%%%%%%%%%%%%%%%%%%%%%%%%%%
\section{Evaluation of the integral with the function $W^{\cal A}$} 
\label{int}
In this appendix we present some details of the derivation of 
Eq.~(\ref{wint}). 
To introduce the integration variable $z$ we exploit the relation
$k_\perp'^2 / (\alpha(1-\alpha)) = M^2 = Q^2(1-z)/z$.
Therefore the $k_\perp'$ integration appearing on the l.h. side of 
Eq.~(\ref{wint}) can be replaced according to
\be
\int \frac{dk_\perp'^2}{\alpha(1-\alpha)} \;\; \longrightarrow \;\;
Q^2 \int^1 \frac{dz}{z^2} \;,
\ee
where just the leading contribution at small $z$ is kept.
Concerning the lower bound of the $z$ integration we refer to the discussion 
given in Sect.~\ref{extg}.

Subsequently, it is convenient to divide the $\alpha$ integration of
Eq.~(\ref{wint}) in two parts by introducing an arbitrary parameter $\mu'$
which fulfills the condition $\Lambda^2 \ll \mu'^2\mst{<}{\sim}Q^2$.
To be specific we separate the symmetric and asymmetric gluon 
configurations in the integration by means of
\be
\int d\alpha = \int_{\mu'^2/M^2}^{1-\mu'^2/M^2} d\alpha
 + \; 2\int_0^{\mu'^2/M^2} d\alpha\,,
\ee
where the symmetric (asymmetric) configurations give rise to the 
first (second) term on the r.h. side of Eq.~(\ref{wint}).
For a given value of $M^2$ the symmetric contribution contains large 
transverse momenta $k_\perp'$ while the asymmetric term is entirely soft.

To extract the leading twist of the symmetric hard part we use
\bea
\int\!\!d^2 k_\perp \frac{N^2\delta_{ij}+2k_i k_j}{N^2+k_\perp^2}\,
\tilde{W}^{\cal A}_{x_\perp}(k_\perp' - k_\perp) \!\!& = &\!\!
C_{ij,a} \int\!\!d^2 k_\perp \, k_a \, 
\tilde{W}^{\cal A}_{x_\perp}(-k_\perp) \; + \; \textrm{higher twist}\,,
\hphantom{mm}
\\
\textrm{with} \quad C_{ij,a} & = & 2\left( \frac{k'_i \delta_{ja} + 
k'_j \delta_{ia}}{N^2+k_\perp'^2} 
- \frac{k'_a (N^2\delta_{ij} + 2k'_i k'_j)}{(N^2+k_\perp'^2)^2} \right) .
\nonumber
\eea
The simple relations
\bea
C_{ij,a}C_{ij,b} & = & \frac{4z(1-z)(1+z^2)}{N^2} \delta_{ab} \,, \;\;
\textrm{and} \label{cmult}
\\
\left| \int d^2 k_\perp k_a \tilde{W}^{\cal A}_{x_\perp}(-k_\perp) \right|^2
& = & (2\pi)^4 \left| \partial_\perp W^{\cal A}_{x_\perp}(0_\perp) \right|^2 
\label{wfourier}
\eea
are important for the further evaluation of the hard contribution. In 
writing Eq.~(\ref{cmult}) we have anticipated the symmetric $k_\perp'$ 
integration and employed the substitution $k'_a k'_b \to k_\perp'^2 
\delta_{ab}/2$. Finally, performing the 
$\alpha$ integration and keeping only the contribution that is dominant 
at small $z$, the first term on the r.h. side of Eq.~(\ref{wint}) is 
obtained. 

In the case of the asymmetric soft contribution, which is independent of 
$Q^2$ and 
therefore leading twist, the $\alpha$ integration is replaced by an 
integral over $k_\perp'$. Since $\alpha$ is small we have $Q^2 \alpha 
\simeq k_\perp'^2 z$, leading to the replacement
\be
\int_0^{\mu'^2/M^2} d\alpha \simeq\frac{z}{Q^2} \int_0^{\mu'^2} 
dk_\perp'^2\,.
\ee
The remaining step to obtain the second term on the r.h. side in 
Eq.~(\ref{wint}) is the $k_\perp$ integration. Keeping in mind that for the 
soft term we can neglect $N^2$ this integration can easily be done with the 
aid of 
\be
\int dk_\perp^2 \, \frac{k_i k_j}{k_\perp^2} \, e^{ik_\perp y_\perp}
 = \frac{2\pi}{y_\perp^2} \left(\delta_{ij} 
   - \frac{2y_i y_j}{y_\perp^2}\right) \,.
\ee

%%%%%%%%%%%%%%%%%%%%%%%%%%%%%%%%%%%%%%%%%%%%%%%%%%%%%%%%%%%%%%%%%%%%%%%%

\section{Simplification of the final expression}\label{simp}
In this appendix, the details of the derivation of Eq.~(\ref{g1ff}), which 
represents a particularly simple form of our final result, are described. 

It is convenient to introduce a real parameter $\epsilon>0$ and to write 
the gluon distribution given by Eqs.~(\ref{fdef}) and (\ref{g1f}) in the 
form 
\be 
xg^{(1)}(x,\mu^2)=\frac{2}{\pi^2}\left(\ln\frac{1}{x}\right)I_0
(0,\mu')\,,
\ee
where
\be
I_\epsilon(a,b)\equiv\int_{a^2<k_\perp'^2<b^2}d^2k_\perp'f_\epsilon
(k_\perp'^2)\,,
\ee
\be
f_\epsilon(k_\perp'^2)\equiv\int\frac{d^2y_\perp}{(2\pi)^2y_\perp^{2-
\epsilon}}\int\frac{d^2z_\perp}{(2\pi)^2z_\perp^{2-\epsilon}}\,h(y_\perp,
z_\perp)\,e^{ik_\perp'(y_\perp-z_\perp)}\left(\frac{2(y_\perp z_\perp)^2}
{y_\perp^2z_\perp^2}-\!1\!\right)\,,
\ee
and
\be
h(y_\perp,z_\perp)\equiv\int d^2x_\perp\mbox{tr}\left[W_{x_\perp}^{\cal A}
(y_\perp)W_{x_\perp}^{\cal A\dagger}(z_\perp)\right]\,. 
\ee 
Furthermore, we have 
\be 
I_0(0,\mu')=\lim_{\epsilon\to 0}\left[I_\epsilon(0,\infty)-
I_\epsilon(\mu',\infty)\right]\,.\label{ielim}
\ee
The $k_\perp'$ integration in the definition of $I_\epsilon(0,\infty)$ is 
easily performed giving a $\delta$-function of $y_\perp-z_\perp$. The 
result is 
\be
I_\epsilon(0,\infty)=\int\frac{d^2y_\perp}{(2\pi)^2y_\perp^{4-2\epsilon}}\,
h(y_\perp,y_\perp)\,.
\ee
Since 
\be
h(y_\perp,z_\perp)\simeq C\,y_\perp\!\cdot\!z_\perp\qquad\mbox{for}\qquad
|y_\perp|\,,|z_\perp|\ll 1/\Lambda\,,
\ee
where $C$ is a constant, one can write 
\be
I_\epsilon(0,\infty)=\frac{C}{2\pi}\,\frac{r^{2\epsilon}}{2\epsilon}+
\frac{1}{4\pi}\int_{r^2<y_\perp^2}\frac{dy_\perp^2}{y_\perp^4}\,
h(y_\perp,y_\perp)+{\cal O}(\epsilon)\,.\label{ie1}
\ee
The parameter $r$ has been introduced to separate the small-distance from 
the large-distance part of the $y_\perp$ integration. 

In the $k_\perp'$ integration defining $I_\epsilon(\mu',\infty)$, the 
momentum variable $k_\perp'$ is always large. Therefore, the result is only 
sensitive to the small distance structure of $h(y_\perp,z_\perp)$, and we 
can write 
\be
I_\epsilon(\mu',\infty)=\int\limits_{\mu'^2<k_\perp'^2}d^2k_\perp'\int
\frac{d^2y_\perp}{(2\pi)^2y_\perp^{2-\epsilon}}\int\frac{d^2z_\perp}
{(2\pi)^2z_\perp^{2-\epsilon}}\,C\,y_\perp\!\cdot\!z_\perp\,e^{ik_\perp'(
y_\perp-z_\perp)}\left(\frac{2(y_\perp z_\perp)^2}{y_\perp^2z_\perp^2}-\!1\!
\right)\,.
\ee
This integral can be calculated using standard methods: 
\be
I_\epsilon(\mu',\infty)=\frac{C}{2\pi^2}\,\frac{\mu'^{-2\epsilon}}
{2\epsilon}\left[\left(\frac{\Gamma(1+\epsilon)\Gamma(
\frac{1-\epsilon}{2})}{\Gamma(1-\frac{\epsilon}{2})}\right)^2(1+\epsilon)
+{\cal O}(\epsilon^2)\right]\,.\label{ie2}
\ee
When inserting Eqs.~(\ref{ie1}) and (\ref{ie2}) in Eq.~(\ref{ielim}), the 
poles in $\epsilon$ cancel. Setting the parameter $r$, which so far was 
arbitrary, to 
\be 
r(\mu)^2=\frac{4e^{1-2\gamma_E}}{\mu'^2}=\frac{4e^{\frac{1}{12}-2\gamma_E}}
{\mu^2}\,,
\ee
the result of Eq.~(\ref{g1ff}) follows. Here $\gamma_E$ is Euler's 
constant. 

%%%%%%%%%%%%%%%%%%%%%%%%%%%%%%%%%%%%%%%%%%%%%%%%%%%%%%%%%%%%%%%%%%%%%%%%
\end{appendix}

\end{document}